\documentclass[paper]{JHEP3}
\usepackage{graphics}
\usepackage{amsmath}

\makeatletter
\@addtoreset{equation}{section}
\makeatother

 \def\unit{\hbox to 3.3pt{\hskip1.3pt \vrule height 7pt width .4pt \hskip.7pt
\vrule height 7.85pt width .4pt \kern-2.4pt
\hrulefill \kern-3pt
\raise 4pt\hbox{\char'40}}}

\def\half{{\textstyle {1 \over 2}}}

\def\ap#1{\alpha^{\prime\,#1}}

\def\slash{\llap /}

\def\makeatletter{\catcode`\@=11}
\makeatletter
\def\mathbox#1{\hbox{$\m@th#1$}}%
\def\math@ccstyles#1#2#3#4#5#6#7{{\leavevmode
      \setbox0\mathbox{#6#7}%
      \setbox2\mathbox{#4#5}%
      \dimen@ #3%
      \baselineskip\z@\lineskiplimit#1\lineskip\z@
      \vbox{\ialign{##\crcr
             \hfil \kern #2\box2 \hfil\crcr
             \noalign{\kern\dimen@}%
             \hfil\box0\hfil\crcr}}}}
\def\mathaccstyles{\math@ccstyles\maxdimen}
\def\maththroughstyles{\math@ccstyles{-\maxdimen}}
\def\unitmatrixDT%
 {\maththroughstyles{.45\ht0}\z@\displaystyle {\mathchar"006C}\displaystyle 1}

\newcommand{\be}{\begin{equation}}
\newcommand{\ee}{\end{equation}}

\newcommand{\bea}{\begin{eqnarray}}
\newcommand{\eea}{\end{eqnarray}}
\newcommand{\nn}{\nonumber}

\newcommand{\beann}{\begin{eqnarray*}}
\newcommand{\eeann}{\end{eqnarray*}}

\newcommand{\pd}{\partial}      
\renewcommand{\L}{\mathcal{L}}  

\newcommand{\g}{\gamma}
\renewcommand{\d}{\delta}
\newcommand{\e}{\epsilon}

\title{Derivative corrections in 10-dimensional super-Maxwell theory}
\author{Andres Collinucci, Mees de Roo
 and Martijn~G.C.~Eenink\\
Institute for Theoretical Physics\\
   Nijenborgh 4, 9747 AG Groningen,\\
     The Netherlands\\
     E-mail: \email{a.collinucci@phys.rug.nl, m.de.roo@phys.rug.nl,
        m.g.c.eenink@phys.rug.nl }}

\preprint{UG-02/42\\ \hepth{0212012}}

\abstract{We construct the supersymmetric effective action
at order $\ap{4}$ of the abelian open superstring.
It includes the $\ap{4}$ terms in the abelian
Born-Infeld action, and in particular the leading derivative correction
of the form $\partial^4F^4$.  Besides linear supersymmetry
this sector of the open string effective action also has a
nonlinear supersymmetry. The terms $\partial^4F^4$ and their
fermionic partners have an arbitrary coefficient, and we discuss
the possible fate of such coefficients when higher orders in
$\ap{}$ are included.}

\keywords{Superstrings and Heterotic Strings, D-branes, Supersymmetric
Effective Theories}
\begin{document}
\section{Introduction\label{Intro}}

The tree-level effective action of the open string, with or without
 Chan-Paton factors, has drawn a lot attention recently \cite{Tseyt1}. Without
 Chan-Paton factors it corresponds, for slowly varying fields, to the
 Born-Infeld action \cite{Fradkin}. Its supersymmetric completion can be obtained
 quite elegantly using $\kappa$-symmetry \cite{APS}-\cite{CGNSW}.
 With Chan-Paton factors the action is known only for some
 low orders of $\ap{}$ \cite{goteborg}-\cite{KS3}. The complications in this
 case are partly due to the fact that for a nonabelian gauge theory
 $[{\cal D},{\cal D}]F = [F,F]$, so that the approximation in which
 derivatives of the fields are ignored can no longer be made. That such
 derivative terms are also present in the abelian case
 is clear from string scattering amplitudes, e.g. the
 four-point function, as mentioned
 in the early papers on the open string effective action
 \cite{Fradkin,GW}. Nevertheless not much is known about
 this extension of the Born-Infeld action at the present time.

In this paper we will investigate these higher
 derivative terms in the context of supersymmetry. The motivation
 for this is, besides the intrinsic interest in the string effective
 action, that these terms also have a relation with the
 nonabelian extension of the Born-Infeld action. One way
 to approach this problem is by using $\kappa$-symmetry,
 in which linear and nonlinear
 supersymmetry arise from the gauge fixing of a local fermionic
 symmetry. Recent efforts in this direction have not been successful
 \cite{BdRS,BBRS}, but alternatives are under investigation
 \cite{Sor}-\cite{PP}. It should be realized that the higher-derivative
 contributions of the {\it abelian} effective action
 do not fit in the present $\kappa$-symmetric formulation
 \cite{APS}-\cite{CGNSW}. Thus a supersymmetric, and eventually a
 $\kappa$-symmetric, formulation of these terms could be helpful
 in solving the nonabelian problem.

In this paper we take a first step in this direction, which is to
 obtain all terms in the abelian effective action through order $\ap{4}$ by
 imposing supersymmetry. The result is that there are two different
 independent supersymmetric invariants.
 The first invariant consists of terms at order $\ap{2}$ and $\ap{4}$,
 respectively of the form $F^4$ and $F^6$ and their fermionic partners, and is the
 contribution to the Born-Infeld invariant through this order. The
 second invariant involves only terms at order $\ap{4}$: $\pd^4F^4$ and their
 fermionic partners. The terms with two derivatives, $\pd^2F^5$ and their
 fermionic counterparts, turn out to be inconsistent with supersymmetry.
 Furthermore, all conceivable terms at $\ap{4}$
 with a higher number of derivatives are removable
 by field redefinitions and therefore trivial.

In discussing the higher derivative contributions to the open string
 effective action it is useful to introduce some notation. We write
 such terms as
 \begin{equation}
 \label{notation}
   {\cal L}_{(m,n)} = {\ap{m}\over g^2} \left(\partial^n F^{p}
        + \partial^{n+1}F^{p-2}\bar\chi\g\chi \right)\,,
 \end{equation}
 where $g$ is the gauge coupling constant of dimension $-(d-4)/2$.
 Henceforth we will set $g=1$. The powers in (\ref{notation})
 are related by $2p-2m+n-4=0$. We will denote the terms at order
 $\ap{m}$ and with $n$ derivatives by $(m,n)$. In this paper we do
 not consider fermion-dependent contributions beyond the bilinear
 fermion terms in (\ref{notation}).

The Noether procedure we employ to find the supersymmetric deformations
 of the super-Maxwell action is an iterative procedure. At a given order
 in $\ap{}$ it yields a number of apparently independent superinvariants,
 all determined up to a multiplicative constant. For example, the $(4,4)$
 terms we will discuss have an arbitrary coefficient $a_{(4,4)}$ that is not
 fixed by supersymmetry. However, some of these coefficients might be determined
 by pursuing the Noether procedure for higher values of $m$ and $n$.
 One can also use input from string theory. We will limit ourselves to the
 contributions to the effective action that follow from the open string tree-level
 {\it S}-matrix. The tree-level correlation functions that one derives from
 the effective action should reproduce these string amplitudes, which allows
 one to fix the previously undetermined coefficients. In particular, the
 $(m,2m-4)$ terms should reproduce the $\ap{}$-expanded 4-point function.
 For $m$ odd, this expansion contains a coefficient
 $\pi^2\zeta(m-2)$. It is hard to see how the Noether procedure could determine such
 coefficients in the absence of algebraic relations between the values
 of the Riemann zeta-function for odd integer arguments. Therefore these
 terms should all correspond to separate independent superinvariants, as
 we argued in \cite{CdRE}. We will come back to these points in the discussion.

Most of the information on derivative corrections concerns bosonic
 terms only. In \cite{AndTs} it was shown that terms $(m,2)$ vanish
 for all $m$. Our calculation of $(4,2)$ confirms this and extends
 it to the corresponding fermionic terms.
 In \cite{AndTs} the bosonic part of
 $(4,4)$ was constructed explicitly.
 More recently, Wyllard \cite{Wyll} obtained the
 $(m,4)$ terms using the boundary state formalism. Further work
 has been done in \cite{Wyll2,DMS} with the Seiberg-Witten map
 and noncommmutativity. Information about the fermionic contributions
 can in principle be obtained from calculating superstring  scattering
 amplitudes involving external fermions. The required formalism
 can be found in \cite{Schwarz}, and applications using the
 four-point function to orders $\ap{m},\ m\leq 4$ can be found in
 \cite{BBRS,Bilal}. The recent determination of the string five-point
 function and its relation with the nonabelian Born-Infeld action
 \cite{BrMaMe} concerns bosonic terms at order $\ap{3}$ only.

This paper is organized as follows. In Section \ref{method} we
 will briefly discuss our method, the main results on linear
 and nonlinear supersymmetry are given in Section \ref{results}.
 Finally, in Section \ref{conclusions}, we discuss the general
 structure of the web of supersymmetric derivative corrections.

\section{Constructing $\ap{}$ corrections\label{method}}

In this section we  review our method
 of imposing supersymmetry
 order by order in $\ap{}$. A more detailed exposition of this
 iterative (or Noether) procedure can be found in \cite{CdRE}.
 Starting point is the $d=10$, $N=1$ supersymmetric Maxwell
 Lagrangian\footnote{We work in Minkowski spacetime and write
 spacetime indices as lower indices. Our conventions for
 the $\g$-matrices follow \cite{Proey}. $\chi$ is a Majorana-Weyl
 spinor and inert under gauge transformations. $\L_m$ is
 the contribution of order $\ap{m}$ to the effective action. Similarly,
 $\delta_m$ indicate supersymmetry transformations of order $\ap{m}$.
 If we want to indicate the part of $\L_m$ with $n$ derivatives we
 write $\L_{(m,n)}$, similarly for $\delta_{(m,n)}$.}
\bea \label{Maxwell}
  \L_0 & = & -\tfrac{1}{4}F_{ab}F_{ab}+\tfrac{1}{2}\bar{\chi}\pd\slash\chi.
\eea
 The equations of motion are simply
\begin{equation}\label{eom0}
 \pd_aF_{ab}=\pd\slash\chi=0\,,
\end{equation}
supersymmetry is realised linearly on the fields:
\bea
  \d_0 A_a & = & \bar{\e}\g_{a}\chi, \nn\\
 \label{ltrans0}
  \d_0\chi & = & \tfrac{1}{2}F_{ab}\g_{ab}\e.
\eea
 Closure of the supersymmetry algebra requires the fields to be
 on-shell and involves a field dependent gauge transformation of
 the gauge field:
\bea
  {[\d_{0\,\e_1},\d_{0\,\e_2}]}A_a & = & 2\bar{\e}_1\pd\slash\e_2 A_a -
            \pd_a ( 2\bar{\e}_1 A\slash\e_2 ), \nn\\
  {[\d_{0\,\e_1},\d_{0\,\e_2}]}\chi & = & 2\bar{\e}_1\pd\slash\e_2 \chi -
            (\tfrac{7}{8}\bar{\e}_1\g_a\e_2\g_a
    -\tfrac{1}{5!16}\bar{\e}_1\g_{abcde}\e_2\g_{abcde}) \pd\slash\chi.
\eea
 This lowest order action also has a nonlinear supersymmetry:
\bea\label{nltrans0}
  \d_0 A_a   &=& 0, \nn\\
  \d_0 \chi  &=& \eta.
\eea
The iterative procedure consists of two steps. Let the $\L_{k}$ for
 $k<m$ be known.  The first step in obtaining the term $\L_m$
 is to write down all possible terms of order $\ap{m}$,
 i.e., terms that have the correct
 dimension and are Lorentz and gauge invariant. We limit ourselves
 to terms that are at most of quadratic order in the fermions.
 Lagrangians are defined up to total derivatives and field
 redefinitions. The possibility for the latter arises when a term
 is proportional to the lowest order equation of motion
 (\ref{eom0}) for one of
 the fields. If such a term is present in $\L_m$ it can be
 removed by a field redefinition of order $m$.
 The price one pays is that the contributions $\L_n$ with $n>m$ are
 modified. We deal with this ambiguity, at each
 order in $\ap{}$, by not allowing in the
 Lagrangian any terms that are proportional to the order $\ap{0}$
 field equations, or terms that can be rewritten as such by means
 of a partial integration. Furthermore, we determine how the
 remaining terms are related by partial integrations and keep only
 an independent subset. This leaves us with a minimal Ansatz for
 $\L_0$ in which each term has an arbitrary
 coefficient that will be determined in the second step.

The second step is to vary the fields in this Ansatz with
 the supersymmetry transformation rules $\delta_0$. In
 addition we need to vary the lower order terms in the Lagrangian,
 say $\L_k$, $k<m$, with the appropriate transformation rules
 $\delta_{m-k}$; both were
 already obtained in a previous stage of the iterative procedure.
 Having done this, we are left with two types of variations. On
 the one hand there are terms which are proportional to the lowest
 order field equation or that can be rewritten as such using a
 partial integration. On the other hand there are variations
 that cannot be rewritten in this way.
 The first set can be eliminated by new
 transformation rules $\delta_m$ of $\L_0$, the second set
 must be set to zero by solving the resulting equations for the
 unknown coefficients in the Ansatz.

In calculating the new transformation rules at order $\ap{m}$
 one will find that some variations
 may be quadratic in the lowest order equations of
 motion. In that case there is an ambiguity in the choice
 of the new transformations $\delta_m$. Regardless of this choice,
 such variations always give rise to transformation rules that contain a lowest order
 equation of motion. Therefore these terms do not play a role in checking the
 closure of the supersymmetry algebra at order $\ap{m}$. If such
 transformations are applied to some $\L_k$ when constructing an invariant
 at order $m+k$, they give variations that can automatically be supersymmetrized.
 Their contribution to the order $m+k$ transformation rules need not contain
 a lower order equation of motion and therefore these terms are important when
 pursuing the Noether procedure to higher orders. Note however that this last
 issue does not yet play a role at order $\ap{4}$ and should not bother
 us in this paper.

The procedure is applied for both linear and nonlinear supersymmetry.

\section{Results\label{results}}

In the end we are left with all possible deformations of the
 Lagrangian and the supersymmetry tranformation rules at a certain
 order in $\ap{}$, up to field redefinitions. In this Section
 we will give the action and study the algebra of the
 linear and nonlinear supersymmetry transformations. The
 transformation rules themselves are given in Appendix A.

\subsection{Orders $\ap{1}$, $\ap{2}$ and $\ap{3}$\label{ap123}}

It is well known that there are no nontrivial supersymmetric
 deformations of (\ref{Maxwell}) at order $\ap{1}$, i.e. all terms
 allowed by supersymmetry can be removed by field redefinitions.
 In \cite{BRS} the terms at order $\ap{2}$ were obtained by using
 the Noether procedure. For completeness, and since we need these
 results at order $\ap{4}$, we review them here.

Following the steps outlined in the previous section, one first
 writes down an Ansatz for the Lagrangian; these are all terms of
 the form $F^4$, $\pd^2 F^3$, $\pd^4 F^2$, $\pd^6 F$ and fermionic
 partners. All of these terms turn out to be removable by field
 redefinitions, except for $F^4$ and its fermionic counterpart $\pd
 F^2\bar{\chi}\g\chi$: $\L_2 = \L_{(2,0)}$.
 Imposing supersymmetry fixes the
 coefficients in this Ansatz up to one overall multiplicative
 constant. The result is:
 \bea
   \L_2 & = &
   {a_{(2,0)}\ap{2}\over 32}\big\{
   -F_{ab}F_{ab}F_{cd}F_{cd}%
   +4\,F_{ac}F_{bd}F_{ab}F_{cd}%
   \nn\\&&
   -8\,F_{ab}F_{ac}\,\bar{\chi}\g_{b}\pd_{c}\chi%
   -2\,F_{ab}\pd_{a}F_{cd}\,\bar{\chi}\g_{bcd}\chi \big\}.%
 \eea
 Of course, these terms are the same as the ones obtained from the
 Born-Infeld invariant\footnote{Up to a field redefinition.}.
 The supersymmetry
 transformations also receive $\ap{2}$ contributions; they are
 given in the appendix. From the point of view of supersymmetry
 the coefficient $a_{(2,0)}$ is arbitrary. At tree-level
 the string four-point function sets $a_{(2,0)}\sim\pi^2$.

At order $\ap{3}$ there are no supersymmetric
 contributions.
 This might be inferred by taking the abelian
 limit of the results of \cite{KS3,CdRE}.
 However, since it is not obvious that every supersymmetric abelian action
 allows a nonabelian supersymmetric extension, it is important to
 check this directly in the abelian context. This has been done in
 \cite{goteborg2} by superspace methods, and
 we have verified this result
 by an independent calculation using the method of Section \ref{method}.

\subsection{Order $\ap{4}$\label{ap4}}

We now turn to the main topic of this paper: the $\ap{4}$
 contributions. There are three nontrivial sectors in the
 Ansatz: $(4,0)$, $(4,2)$, $(4,4)$.
 The structures with more derivatives are removable
 by field redefinitions. Furthermore, it turns out that this is also the case
 for the bosonic terms $(4,2)$, i.e., the terms $\pd^2 F^5$,
 but not for their fermionic
 partners, which are of the form $\pd^3 F^3 \bar{\chi}\g\chi$.
 In applying the method of Section \ref{method} we need the
 variations $\delta_0\L_{(4,0)}$ as well as
 $\delta_2\L_{(2,0)}$. In the cases $(4,2)$ and $(4,4)$
 only the variation $\delta_0$ is needed.

The results of the Noether procedure are the following: in the
 sector $\L_{(4,0)}$ with $F^6$ and $\pd F^4 \bar{\chi}\g\chi$ the only terms
 allowed by supersymmetry are those needed for the `continuation'
 of the invariant of order $\ap{2}$, i.e. the Born-Infeld invariant.
 Thus there appears no {\it new}
 invariant, independent of the lower orders.
 Furthermore, the fermionic terms in $\L_{(4,2)}$ of the form $\pd^3
 F^3 \bar{\chi}\g\chi$ are not supersymmetrizable.
 Finally, in the
 section $\L_{(4,4)}$ there does appear a new invariant. We have
 verified that the bosonic terms of this invariant are the same as
 those of \cite{Wyll,KS3} up to field redefinitions. We
 however disagree with \cite{Bilal}, where both bosonic and
 fermionic terms are determined by comparison with the
 string four-point function.

The action at order $\ap{4}$, $\L_4 = \L_{(4,0)}+\L_{(4,4)}$,  reads:
 \bea
  \mathcal{L}_{(4,0)} & = &
  {(a_{(2,0)})^2\ap{4}\over 384}\big\{
  -32\,F_{ab}F_{bc}F_{cd}F_{de}F_{ef}F_{af}%
  -12\,F_{ab}F_{bc}F_{cd}F_{ad}F_{ef}F_{ef}%
  \nn\\&&
  -\,F_{ab}F_{ab}F_{cd}F_{cd}F_{ef}F_{ef}%
  -12\,\pd_{a}F_{bc}F_{de}F_{af}F_{be}\,\bar{\chi}\g_{cdf}\chi%
  \nn\\&&
  +72\,F_{ab}F_{cd}F_{be}F_{de}\,\bar{\chi}\g_{a}\pd_{c}\chi%
  +18\,\pd_{a}F_{bc}F_{de}F_{ef}F_{af}\,\bar{\chi}\g_{bcd}\chi%
  \nn\\&&
\label{L40}
  +12\,\pd_{a}F_{bc}F_{de}F_{bf}F_{ae}\,\bar{\chi}\g_{cdf}\chi \big\} \,,
  \nn\\
  &&
  \nn\\
  \L_{(4,4)} & = &
  a_{(4,4)} \ap{4}\big\{
  -8\,F_{ab}F_{bc}\pd_{d}\pd_{e}F_{af}\pd_{d}\pd_{e}F_{cf}%
  -8\,F_{ab}\pd_{c}F_{ad}\pd_{e}F_{bf}\pd_{c}\pd_{e}F_{df}%
  \nn\\&&
  +32\,F_{ab}\pd_{c}F_{ad}\pd_{e}F_{bf}\pd_{d}\pd_{e}F_{cf}%
  +16\,F_{ab}\pd_{c}F_{de}\pd_{a}F_{ef}\pd_{d}\pd_{f}F_{bc}%
  \nn\\&&
  +4\,\pd_{a}\pd_{b}F_{cd}\pd_{a}\pd_{b}F_{ce}\,\bar{\chi}\g_{d}\pd_{e}\chi%
  -4\,\pd_{a}F_{bc}\pd_{a}\pd_{d}F_{ef}\,\bar{\chi}\g_{bef}\pd_{c}\pd_{d}\chi%
  \nn\\&&
  +4\,F_{ab}\pd_{c}\pd_{d}F_{ef}\,\bar{\chi}\g_{abe}\pd_{c}\pd_{d}\pd_{f}\chi%
  +8\,F_{ab}\pd_{c}\pd_{d}F_{ae}\,\bar{\chi}\g_{b}\pd_{c}\pd_{d}\pd_{e}\chi%
  \nn\\&&
\label{L44}
  +2\,\pd_{a}F_{bc}\pd_{a}\pd_{d}\pd_{e}F_{bc}\,\bar{\chi}\g_{d}\pd_{e}\chi \big\}%
 \eea
Note that the overall coefficient of $\L_{(4,0)}$ is uniquely fixed by
 supersymmetry in terms of $a_{(2,0)}$, the coefficient $a_{(4,4)}$ is unrelated.
 String theory tells us that at tree-level $a_{(4,4)}\sim\pi^4$.

\subsection{Nonlinearly realised SUSY\label{nlsusy}}

Nonlinear supersymmetry arises from the breaking of $N=2$ supersymmetry
 to $N=1$: the superstring effective action
 corresponds to the worldvolume theory of a D9-brane, and D-branes
 break half of the $N=2$ supersymmetry. In the $\kappa$-symmetric
 formulation of the Born-Infeld action it arises from the gauge-fixing
 of the local $\kappa$-symmetry. The presence of this nonlinear symmetry can
 be taken as an indication that a $\kappa$-symmetric formulation is possible,
 but there is certainly no proof of such a relation.

The nonlinear supersymmetry is quite restrictive in the sectors without
 extra derivatives: $(2,0)$ and $(4,0)$. For instance, in $(4,0)$ one
 finds the result $\L_{(4,0)}$ (\ref{L40}), plus one additional term which
 is invariant under the nonlinear symmetry (up to variations which vanish
 on-shell) all by itself. It is to be noted that in the sector $(4,2)$
 it is possible to impose nonlinear supersymmetry. Similarly, in $(4,4)$
 nonlinear supersymmetry is not restrictive at all, and many combinations
 of terms are invariant under (\ref{nltrans0}).

\subsection{Closure of the algebra\label{algebra}}

The supersymmetry algebra can only be evaluated on the vector field, due to
 the absence of higher fermion terms in the action and transformation rules.
 We find, for all contributions of order $\ap{m},\ m\leq 4$:
\bea
  [\d_{\e_1},\d_{\e_2}]A_a & = &
  2\bar{\e}_1\pd\slash\e_2 A_a - \pd_a(2\bar{\e}_1A\slash\e_2) +
  \nn\\&&
  + a_{(4,4)}\,\pd_a\big\{
  +32\,\pd_{c}\pd_{d}F_{eb}\pd_{c}F_{df}F_{ef}\,\bar{\e}_1\g_{b}\e_2%
  +16\,\pd_{c}\pd_{d}F_{eb}\pd_{c}F_{ef}F_{df}\,\bar{\e}_1\g_{b}\e_2%
  \nn\\&&
  +8\,\pd_{c}\pd_{d}F_{ef}\pd_{c}F_{ef}F_{db}\,\bar{\e}_1\g_{b}\e_2%
  +16\,\pd_{c}F_{db}\pd_{e}F_{fc}\pd_{e}F_{fd}\,\bar{\e}_1\g_{b}\e_2%
  \nn\\&&
  -16\,\pd_{c}F_{db}\pd_{e}F_{fc}\pd_{f}F_{ed}\,\bar{\e}_1\g_{b}\e_2%
  +4\,\pd_{g}\pd_{h}F_{bc}\pd_{g}F_{de}F_{hf}\,\bar{\e}_1\g_{bcdef}\e_2%
  \nn\\&&
  +4\,\pd_{g}\pd_{h}F_{bc}\pd_{g}F_{hd}F_{ef}\,\bar{\e}_1\g_{bcdef}\e_2%
  -4\,\pd_{g}F_{bc}\pd_{h}F_{de}\pd_{g}F_{hf}\,\bar{\e}_1\g_{bcdef}\e_2
  \big\}\,.
\eea
Note that the terms without extra derivatives do not modify the gauge
transformation in the algebra. This is simply due to the absence of
a derivative in the corresponding transformation rules. Nevertheless, the
required cancellations for closure, and the fact that all remaining
terms combine into a gauge transformation, is an important check on our
result.

The algebra of the nonlinear transformations reads \cite{goteborg}:
\bea
 [\d_{\eta_1},\d_{\eta_2}]A_a & = &
    {a_{(2,0)}\ap{2}\over 2}\left(  \bar\eta_1\pd\slash\eta_2 A_a  -
             \partial_a ( \bar\eta_1 A \slash\eta_2 ) \right)\,,
\eea
which does not have modifications at order $\ap{4}$. Note
that this is just the usual supersymmetry algebra, occurring at a higher order in $\ap{}$.
This {\it proves} that the nonlinear symmetry is indeed a supersymmetry.

The mixed algebra takes on the form:
\bea
 [\d_{\e},\d_{\eta}]A_a & = &
    {a_{(4,4)}\ap{2}\over 80 }\,\partial_a \big(
       \partial_b\partial_cF_{de}
       (5\partial_cF_{de} \bar\e\gamma_b\eta
       +2\partial_bF_{cf} \bar\e\gamma_{def}\eta)\big) \,.
\eea

\section{Discussion and conclusions\label{conclusions}}

\begin{figure} \label{invariants.eps}
  \begin{center}
    \scalebox{1.00}{\includegraphics{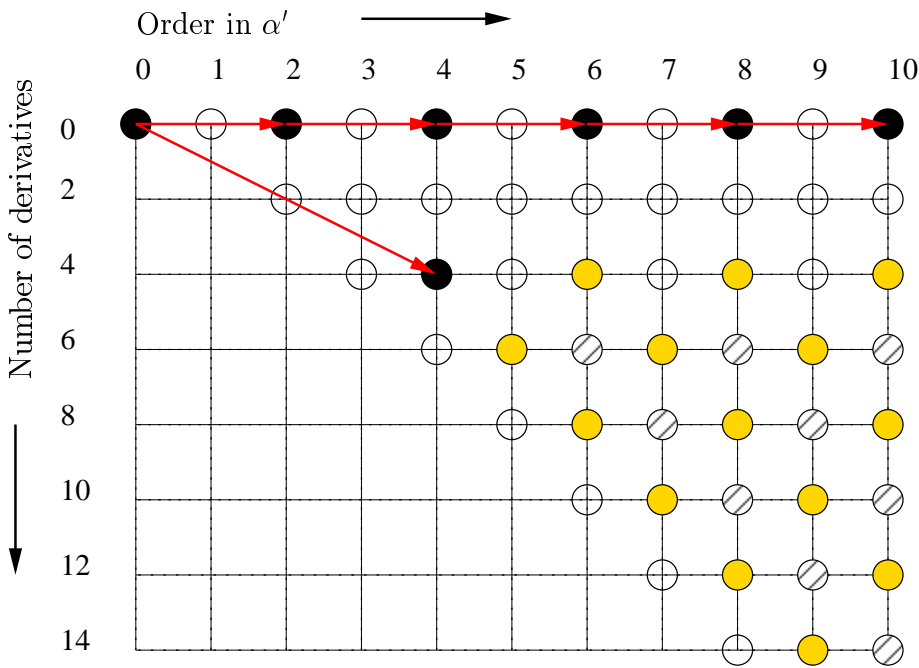}}
  \end{center}
  \caption{Structure of the abelian open superstring tree level effective
 action. Black dots indicate nonempty sectors of which the explicit form is known.
 Empty white dots correspond to sectors that
 are known to be empty up to field redefinitions, already taking into account
 conjecture 1. Yellow dots indicate sectors that are known to be nonempty, but
 have yet to be constructed explicitly. Slashed white dots indicate sectors that
 should be empty by conjecture 2. The red arrows indicate the known supersymmetry
 transformation rules.}
\end{figure}

Our present knowledge of the open string effective action is
 represented in Figure 1. Black dots indicate the sectors $(m,n)$
 for which both bosonic and fermionic terms have been
 established. The dots $(m,0)$
 for all $m$ form the Born-Infeld action, for which the result is
 known to all orders in the fermions \cite{APS}-\cite{CGNSW}; for the single
 dot $(4,4)$, presented in this paper,
 only the terms bilinear in the fermions are known. White dots
 without diagonal lines
 are known to be empty. These include all points $(m,2(m-1))$,
 that would correspond to the three-point function\footnote{In addition,
 it turns out that all terms that one could possibly write down in this
 sector can be removed by field redefinitions.}. We have also
 put a white dot for all points $(m,2)$, although strictly speaking
 the absence of these contributions has only been established
 for the bosonic terms. As far as we know there is, in this order
 by order superinvariant, no way to exclude a priori the presence of a
 fermionic contribution in these sectors. Nevertheless,
 we make here the first conjecture:

\noindent{\bf Conjecture 1:}\ If the bosonic part of the sector
 $(m,n)$ vanishes, then also the fermionic contributions of
 that sector vanish.

\noindent The white dots also include all points
 $(m,4)$ for $m$ odd. In \cite{Wyll} the bosonic part of $(m,4)$ was
 obtained for all $m$; it vanishes for $m$ odd. In favor of this
 conjecture is our result for $(4,2)$. The conjecture implies that
 the terms $(m,0)$, $m$ odd, all vanish. This is obvious for the
 bosonic terms, but it has to be checked for the fermionic terms.
 This could be done for instance by starting with the
 explicit form of the Born-Infeld
 action given by \cite{APS} and by doing the field redefinitions
 needed to eliminate all fermionic contributions at odd orders in $\ap{}$.
 For the higher derivative terms a useful check would be $(5,4)$.

There are many sectors which are known to be present but
 for which the complete contribution to the effective action
 is presently unknown. These yellow dots
 include all points $(m,2m-4)$
 corresponding to the four-point function, all points
 $(m,4)$, $m$ even \cite{Wyll},
 and we would add to these all points $(m,2m-4k+4)$
 (for $m>2k-1$ to avoid $n=2$)
 which correspond to higher derivative
 contributions to the open string $2k$-point function.

The remaining points are white with diagonal lines, and
 correspond to contributions to the string $2k+1$-point functions
 for $k>1$. Everything we know so far about the effective action
 would be consistent with the vanishing of these contributions,
 which would imply the vanishing of the
 open string $2k+1$-point function without Chan-Paton factors.
 Note that present knowledge confirms this conjecture
 for terms with 0, 2 and 4 extra derivatives. Thus we make a second
 conjecture\footnote{It is  a pleasure to thank A.~Tseytlin for
 an interesting correspondence on this point.}:

\noindent{\bf Conjecture 2:}\ The tree level open string
 {\it odd}-point function without Chan-Paton factors vanishes.

\noindent Since this is a conjecture about string theory, it
 cannot be checked by using supersymmetry alone. For the
 effective action it implies that there are no terms
 with an odd number of fields (either bosons or fermions).

With these conjectures in mind we analyze the supersymmetry
 transformations that connect the dots in Figure 1. We have drawn
 arrows to indicate the known supersymmetry transformations. As a
 first example we consider the terms $(m,0)$, $m$ even, i.e., the Born-Infeld
 invariant. These terms are invariant under the transformations
 $\d_0, \d_{(2,0)}, \d_{(4,0)}, \ldots$, depending only on the single
 parameter $a_{(2,0)}$. Note that we indicate these transformations by
 a repeated addition of the same arrow, and {\it not} by drawing new arrows from
 $(0,0)$ to $(m,0)$ for each $m$. In this way we denote that all these terms
 contribute to the same invariant. Similarly, the point $(4,4)$ is the
 leading term in a new sequence of supersymmetry transformations that continues
 to the points $(4k,4k)$, involving the parameter $a_{(4,4)}$. It is clear
 that all points on the diagonal $(m,2m-4)$ will lead to
 at least one new sequence of arrows
 or supersymmetry transformations, involving parameters $a_{(m,2m-4)}$. The question
 is now, whether these new `independent' invariants will remain independent when
 the Noether procedure is pursued to higher orders. Consider for example the
 point $(8,8)$. This point can be reached from $(0,0)$
 by applying the arrow $(4,4)$ twice, but also
 by applying the arrow $(2,0)$ and then $(6,8)$ (or vice versa). These
 contributions need to be cancelled by
 the $\delta_0$ variation of ${\cal L}_{(8,8)}$. In principle,
 there are now two possibilities: they can either be cancelled
 separately, or not. In the latter case we need both contributions at the same
 time, and then there must be
 relations between the coefficients $a_{(4,4)}a_{(4,4)}$ and
 $a_{(2,0)}a_{(6,8)}$. So it is indeed possible that a priori independent invariants
 are related to each other at higher orders in the iteration.
 Note however that at least $a_{(2,0)}$ and $a_{(4,4)}$
 will remain independent to all orders. The reason is that $a_{(2,0)}$
 and $a_{(4,4)}$ can be changed independently by rescaling $\ap{}$ and the
 extra derivatives, respectively.

Now we invoke our knowledge of the string tree-level 4-point function \cite{CdRE}.
 In the tree-level effective action the coefficients $a_{(m,2m-4)}$ for $m$ even
 contain factors $\pi^m$, and therefore the coefficients $a_{(2,0)},\ a_{(4,4)}$
 and $a_{(6,8)}$ are, in string theory, all proportional to powers of $\pi$.
 The relations between these coefficients alluded to in the previous paragraph
 are therefore possible in the string theory context. Independence  of these
 coefficients would imply that there are more supersymmetric
 invariants than required by string theory.

At odd $m$, and also at even $m$ for $m$ large enough, the situation is different.
 At odd $m$ the string four-point function has a factor
 $\pi^2\zeta(m-2)$. This implies that the coefficients $a_{(m,2m-4)}$,
 for $m$ odd, will remain independent to all orders in
 the Noether procedure, since there are no (known) relations between the values of
 the Riemann $\zeta$-function for odd values of its argument.
 At certain even values of $m$ one finds, besides powers of $\pi$, also
 terms with $\zeta$-functions.
 An interesting example is the point $(10,12)$. We can reach it with
 supersymmetry transformations through $(6,8)$ by applying $\delta_{(4,4)}$,
 from $(8,12)$ by applying
 $\delta_{(2,0)}$,  and from $(5,6)$ by applying $\delta_{(5,6)}$.
 Now it should be noted that at $(8,12)$ the
 string four-point function has two separate kinematic structures: one
 proportional to $\pi^{10}$, and one proportional to $\pi^2\zeta(3)^2$.
 The three different ways of arriving at $(10,12)$ therefore lead to
 two kinds of terms: those with $\pi^{12}$ and those with $\pi^4\zeta(3)^2$.
 These must belong to two separate supersymmetric invariants. Again, the
 minimum requirement is that these are part of the invariants containing
 $(4,4)$ and $(5,6)$, anything else leads to an accumulation of more and more
 invariants not required by string theory.

The minimum assumption is therefore that supersymmetry requires independent
 coefficients at $(4,4)$, and at $(m,2m-4)$ with $m$ odd, the points where
 $\pi^2\zeta(m-2)$ appears. This leads to the conjecture:

\noindent{\bf Conjecture 3:}\ The sectors $\L_{(4,4)}$ and
 $\L_{(m,2m-4)}$, $m$ odd, contain the
 leading contributions to separate superinvariants. There are no other
 invariants starting at  $\L_{(m,n)}$ for any $m,\ n$.

The independent coefficients in  the maximal extension of supersymmetric
 Maxwell theory in $d=10$ are, according to these conjectures, $a_{(2,0)}$,
 $a_{(4,4)}$ and $a_{(m,2m-4)}$ for $m$ odd. The tree-level open string
 effective action corresponds to a particular choice of these coefficients.
 The independence of $a_{(2,0)}$ and $a_{(4,4)}$ implies that
 the Born-Infeld action for slowly varying fields is a separate invariant.

The issues that we raised above clearly need to be addressed.
 It is probably not possible to continue the Noether procedure we used much further,
 due to the rapidly increasing number of possible terms in the Lagrangian at higher orders
 in $\ap{}$. One should therefore look for other methods of tackling these issues.
 In particular, it would be interesting to see whether more information on
 the structure of the superinvariants can be obtained from string theory
 considerations. Another possibility would be to set up the Noether procedure in
 $d=10$ $\mathcal{N}=1$ on-shell superspace. A clear advantage of this setting is that
 field redefinition ambiguities do not arise, since all fields are constrained to satisfy
 their lowest order equations of motion. Finally, the persistence of the non-linear
 supersymmetry in the higher-derivative terms  is a strong indication that a
 $\kappa$-symmetric formulation of the all-order effective action exists. Given the
 success of $\kappa$-symmetry in clarifying the structure of the supersymmetric
 Born-Infeld action, it is conceivable, if not likely, that it will
 yield similar striking results when applied to this problem.

\section*{Acknowledgements}

We thank Arkady Tseytlin for an interesting correspondence.
The work of Martijn Eenink is part of the research
programme of the ``Stichting
voor Fundamenteel Onderzoek van de Materie'' (FOM).
This work is supported in part by the European
Commission RTN programme HPRN-CT-2000-00131, in which we
are associated to the University of Utrecht.

\appendix

\section{Transformation rules}

In the Appendix we will give the complete set of transformation rules
for all orders $\ap{m},\ m\leq 4$. As discussed in Section \ref{method},
we give only those transformations that do not vanish on-shell.
These are needed
to establish the existence of an invariant, but do not change the
supersymmetry algebra.

\subsection{Linear supersymmetry\label{lintrans}}

The variation of the gauge field reads:
\bea
  \d A_a & = & \bar\e\gamma_a\chi
  \nn\\&&
  + {a_{(2,0)}\ap{2}\over 32}\big\{
  -6\,F_{cd}F_{cd}\,\bar{\e}\g_{a}\chi%
  -16\,F_{ac}F_{cd}\,\bar{\e}\g_{d}\chi%
  \nn\\&&
  -4\,F_{ac}F_{de}\,\bar{\e}\g_{cde}\chi%
  +\,F_{cd}F_{ef}\,\bar{\e}\g_{acdef}\chi \big\}%
  \nn\\&&
  \nn\\&&
  + {a_{(2,0)}^2\ap{4}\over 1024} \big\{
  +96\,F_{ab}F_{cd}F_{cd}F_{be}\,\bar{\e}\g_{e}\chi%
  -128\,F_{ab}F_{bc}F_{cd}F_{de}\,\bar{\e}\g_{e}\chi%
  \nn\\&&
  +104\,F_{bc}F_{bd}F_{ce}F_{de}\,\bar{\e}\g_{a}\chi%
  +18\,F_{bc}F_{bc}F_{de}F_{de}\,\bar{\e}\g_{a}\chi%
  \nn\\&&
  +32\,F_{ab}F_{cd}F_{ce}F_{df}\,\bar{\e}\g_{bef}\chi%
  +24\,F_{ab}F_{cd}F_{cd}F_{ef}\,\bar{\e}\g_{bef}\chi%
  \nn\\&&
  +32\,F_{ab}F_{bc}F_{de}F_{cf}\,\bar{\e}\g_{def}\chi%
  -16\,F_{ab}F_{cd}F_{be}F_{fg}\,\bar{\e}\g_{cdefg}\chi%
  \nn\\&&
  -6\,F_{bc}F_{bc}F_{de}F_{fg}\,\bar{\e}\g_{adefg}\chi%
  -16\,F_{bc}F_{bd}F_{ce}F_{fg}\,\bar{\e}\g_{adefg}\chi%
  \nn\\&&
  -\tfrac{4}{3}F_{ab}F_{cd}F_{ef}F_{gh}\,\bar{\e}\g_{bcdefgh}\chi%
  +\tfrac{1}{6}F_{bc}F_{de}F_{fg}F_{hi}\,\bar{\e}\g_{abcdefghi}\chi \big\} + %
  \nn
\eea
\bea
  &&
 +\ a_{(4,4)}\ap{4}\big\{
  +20\,\pd_{b}F_{cd}\pd_{a}\pd_{b}\pd_{e}F_{cd}\,\bar{\e}\g_{e}\chi%
  +28\,\pd_{b}\pd_{c}F_{ad}\pd_{b}\pd_{d}F_{ce}\,\bar{\e}\g_{e}\chi%
  \nn\\&&
  -28\,\pd_{b}\pd_{c}F_{ad}\pd_{b}\pd_{c}F_{de}\,\bar{\e}\g_{e}\chi%
  +8\,\pd_{b}F_{cd}\pd_{a}\pd_{b}\pd_{c}F_{ef}\,\bar{\e}\g_{def}\chi%
  \nn\\&&
  +2\,\pd_{a}\pd_{b}F_{cd}\pd_{b}\pd_{c}F_{ef}\,\bar{\e}\g_{def}\chi%
  +4\,\pd_{b}\pd_{c}F_{de}\pd_{b}\pd_{c}F_{fg}\,\bar{\e}\g_{adefg}\chi%
  \nn\\&&
  +4\,F_{bc}\pd_{a}\pd_{d}\pd_{e}F_{bc}\,\bar{\e}\g_{d}\pd_{e}\chi%
  +12\,\pd_{b}F_{cd}\pd_{b}\pd_{e}F_{cd}\,\bar{\e}\g_{a}\pd_{e}\chi%
  \nn\\&&
  +4\,\pd_{b}F_{cd}\pd_{b}\pd_{e}F_{cd}\,\bar{\e}\g_{e}\pd_{a}\chi%
  -84\,\pd_{b}F_{cd}\pd_{a}\pd_{c}F_{be}\,\bar{\e}\g_{d}\pd_{e}\chi%
  \nn\\&&
  +72\,\pd_{b}F_{cd}\pd_{a}\pd_{b}F_{ce}\,\bar{\e}\g_{d}\pd_{e}\chi%
  -40\,\pd_{b}F_{cd}\pd_{b}\pd_{c}F_{ae}\,\bar{\e}\g_{d}\pd_{e}\chi%
  \nn\\&&
  +8\,\pd_{b}F_{cd}\pd_{a}\pd_{c}F_{be}\,\bar{\e}\g_{e}\pd_{d}\chi%
  +40\,\pd_{b}F_{cd}\pd_{b}\pd_{c}F_{ae}\,\bar{\e}\g_{e}\pd_{d}\chi%
  \nn\\&&
  +8\,\pd_{b}F_{cd}\pd_{a}\pd_{b}F_{ce}\,\bar{\e}\g_{e}\pd_{d}\chi%
  -8\,\pd_{b}F_{ac}\pd_{b}\pd_{c}F_{de}\,\bar{\e}\g_{d}\pd_{e}\chi%
  \nn\\&&
  +16\,\pd_{b}F_{ac}\pd_{b}\pd_{d}F_{ce}\,\bar{\e}\g_{d}\pd_{e}\chi%
  -8\,\pd_{b}F_{ac}\pd_{c}\pd_{d}F_{be}\,\bar{\e}\g_{d}\pd_{e}\chi%
  \nn\\&&
  +4\,F_{bc}\pd_{a}\pd_{b}\pd_{d}F_{ef}\,\bar{\e}\g_{cef}\pd_{d}\chi%
  -8\,\pd_{b}F_{cd}\pd_{b}\pd_{c}F_{ef}\,\bar{\e}\g_{ade}\pd_{f}\chi%
  \nn\\&&
  +8\,\pd_{b}F_{cd}\pd_{c}\pd_{e}F_{bf}\,\bar{\e}\g_{ade}\pd_{f}\chi%
  +4\,\pd_{b}F_{cd}\pd_{b}\pd_{c}F_{ef}\,\bar{\e}\g_{aef}\pd_{d}\chi%
  \nn\\&&
  +6\,\pd_{b}F_{cd}\pd_{a}\pd_{b}F_{ef}\,\bar{\e}\g_{cde}\pd_{f}\chi%
  +4\,\pd_{b}F_{cd}\pd_{a}\pd_{b}F_{ef}\,\bar{\e}\g_{cef}\pd_{d}\chi%
  \nn\\&&
  +2\,\pd_{b}F_{cd}\pd_{a}\pd_{e}F_{bf}\,\bar{\e}\g_{cdf}\pd_{e}\chi%
  +4\,\pd_{b}F_{cd}\pd_{b}\pd_{e}F_{fg}\,\bar{\e}\g_{acdfg}\pd_{e}\chi%
  \nn\\&&
  +16\,F_{bc}\pd_{b}\pd_{d}F_{ae}\,\bar{\e}\g_{e}\pd_{c}\pd_{d}\chi%
  -56\,F_{bc}\pd_{a}\pd_{d}F_{be}\,\bar{\e}\g_{e}\pd_{c}\pd_{d}\chi%
  \nn\\&&
  +24\,F_{ab}\pd_{b}\pd_{c}F_{de}\,\bar{\e}\g_{d}\pd_{c}\pd_{e}\chi%
  -80\,F_{bc}\pd_{a}\pd_{b}F_{de}\,\bar{\e}\g_{d}\pd_{c}\pd_{e}\chi%
  \nn\\&&
  -4\,F_{bc}\pd_{d}\pd_{e}F_{bc}\,\bar{\e}\g_{a}\pd_{d}\pd_{e}\chi%
  +84\,F_{bc}\pd_{b}\pd_{d}F_{ae}\,\bar{\e}\g_{c}\pd_{d}\pd_{e}\chi%
  \nn\\&&
  -116\,F_{bc}\pd_{d}\pd_{e}F_{ab}\,\bar{\e}\g_{c}\pd_{d}\pd_{e}\chi%
  +24\,\pd_{b}F_{cd}\pd_{b}F_{ce}\,\bar{\e}\g_{a}\pd_{d}\pd_{e}\chi%
  \nn\\&&
  -16\,\pd_{b}F_{cd}\pd_{c}F_{be}\,\bar{\e}\g_{a}\pd_{d}\pd_{e}\chi%
  +8\,\pd_{b}F_{cd}\pd_{b}F_{ce}\,\bar{\e}\g_{e}\pd_{a}\pd_{d}\chi%
  \nn\\&&
  -8\,\pd_{b}F_{cd}\pd_{c}F_{be}\,\bar{\e}\g_{e}\pd_{a}\pd_{d}\chi%
  +64\,\pd_{b}F_{cd}\pd_{a}F_{ce}\,\bar{\e}\g_{e}\pd_{b}\pd_{d}\chi%
  \nn\\&&
  +8\,\pd_{b}F_{cd}\pd_{c}F_{ae}\,\bar{\e}\g_{e}\pd_{b}\pd_{d}\chi%
  -16\,\pd_{b}F_{ac}\pd_{b}F_{de}\,\bar{\e}\g_{d}\pd_{c}\pd_{e}\chi%
  \nn\\&&
  +56\,\pd_{b}F_{ac}\pd_{c}F_{de}\,\bar{\e}\g_{d}\pd_{b}\pd_{e}\chi%
  -16\,\pd_{b}F_{ac}\pd_{d}F_{be}\,\bar{\e}\g_{e}\pd_{c}\pd_{d}\chi%
  \nn\\&&
  -8\,\pd_{b}F_{ac}\pd_{d}F_{ce}\,\bar{\e}\g_{e}\pd_{b}\pd_{d}\chi%
  -4\,F_{ab}\pd_{c}\pd_{d}F_{ef}\,\bar{\e}\g_{bef}\pd_{c}\pd_{d}\chi%
  \nn\\&&
  -2\,F_{bc}\pd_{a}\pd_{d}F_{ef}\,\bar{\e}\g_{bce}\pd_{d}\pd_{f}\chi%
  -8\,F_{bc}\pd_{b}\pd_{d}F_{ef}\,\bar{\e}\g_{ace}\pd_{d}\pd_{f}\chi%
  \nn\\&&
  -8\,F_{bc}\pd_{d}\pd_{e}F_{bf}\,\bar{\e}\g_{acf}\pd_{d}\pd_{e}\chi%
  -4\,F_{bc}\pd_{a}\pd_{d}F_{ef}\,\bar{\e}\g_{bef}\pd_{c}\pd_{d}\chi%
  \nn\\&&
  -4\,\pd_{a}F_{bc}\pd_{d}F_{ef}\,\bar{\e}\g_{bce}\pd_{d}\pd_{f}\chi%
  -4\,\pd_{a}F_{bc}\pd_{d}F_{ef}\,\bar{\e}\g_{bef}\pd_{c}\pd_{d}\chi%
  \nn\\&&
  +4\,\pd_{b}F_{cd}\pd_{b}F_{ef}\,\bar{\e}\g_{cde}\pd_{a}\pd_{f}\chi%
  +4\,\pd_{b}F_{cd}\pd_{e}F_{bf}\,\bar{\e}\g_{acd}\pd_{e}\pd_{f}\chi%
  \nn\\&&
  +8\,\pd_{b}F_{cd}\pd_{c}F_{ef}\,\bar{\e}\g_{ade}\pd_{b}\pd_{f}\chi%
  +4\,F_{bc}\pd_{d}\pd_{e}F_{fg}\,\bar{\e}\g_{abcfg}\pd_{d}\pd_{e}\chi%
  \nn\\&&
  +2\,\pd_{b}F_{cd}\pd_{e}F_{fg}\,\bar{\e}\g_{acdfg}\pd_{b}\pd_{e}\chi%
  +16\,F_{ab}\pd_{c}F_{de}\,\bar{\e}\g_{d}\pd_{b}\pd_{c}\pd_{e}\chi%
  \nn\\&&
  +32\,F_{bc}\pd_{d}F_{be}\,\bar{\e}\g_{a}\pd_{c}\pd_{d}\pd_{e}\chi%
  +4\,F_{bc}\pd_{d}F_{ef}\,\bar{\e}\g_{bce}\pd_{a}\pd_{d}\pd_{f}\chi \big\}\,.%
\eea

The variations of the fermion are:
\bea
  \d\chi & = &  \half\gamma_{ab}\e F_{ab}
  \nn\\&&
  +{a_{(2,0)}\ap{2}\over 32}\big\{
  +F_{ab}F_{cd}F_{cd}\,\g_{ab}\e%
  -4\,F_{ab}F_{cd}F_{ac}\,\g_{bd}\e%
  \nn\\&&
  +\tfrac{1}{6}\,F_{ab}F_{cd}F_{ef}\,\g_{abcdef}\e \big\}+%
  \nn
\eea
\bea
  &&
  +{a_{(2,0)}^2\ap{2}\over 1024} \big\{
  -12\,F_{ab}F_{ac}F_{bd}F_{cd}F_{ef}\,\g_{ef}\e%
  +F_{ab}F_{ab}F_{cd}F_{cd}F_{ef}\,\g_{ef}\e%
  \nn\\&&
  -8\,F_{ab}F_{ab}F_{cd}F_{ce}F_{df}\,\g_{ef}\e%
  +64\,F_{ab}F_{ac}F_{bd}F_{ce}F_{df}\,\g_{ef}\e%
  \nn\\&&
  +\tfrac{1}{3}\,F_{ab}F_{ab}F_{cd}F_{ef}F_{gh}\,\g_{cdefgh}\e%
  -4\,F_{ab}F_{cd}F_{ae}F_{bf}F_{gh}\,\g_{cdefgh}\e%
  \nn\\&&
  +\tfrac{1}{60}\,F_{ab}F_{cd}F_{ef}F_{gh}F_{ij}\,\g_{abcdefghij}\e \big\} +%
  \nn\\&&
  \nn\\&&
  +a_{(4,4)}\ap{4}\big\{
  -8\,\g_{ab}\e\, F_{ac}\pd_{d}F_{ef}\pd_{b}\pd_{c}\pd_{d}F_{ef}%
  -16\,\g_{ab}\e\, F_{cd}\pd_{e}F_{cf}\pd_{d}\pd_{e}\pd_{f}F_{ab}%
  \nn\\&&
  -8\,\e\,\pd_{a}F_{bc}\pd_{b}F_{de}\pd_{a}\pd_{c}F_{de}%
  +2\,\g_{ab}\e\, F_{cd}\pd_{e}\pd_{f}F_{ab}\pd_{e}\pd_{f}F_{cd}%
  \nn\\&&
  +72\,\g_{ab}\e\, F_{cd}\pd_{c}\pd_{e}F_{af}\pd_{e}\pd_{f}F_{bd}%
  -40\,\g_{ab}\e\, F_{cd}\pd_{c}\pd_{e}F_{af}\pd_{d}\pd_{f}F_{be}%
  \nn\\&&
  +16\,\g_{ab}\e\, F_{cd}\pd_{c}\pd_{e}F_{af}\pd_{d}\pd_{e}F_{bf}%
  -56\,\g_{ab}\e\, F_{cd}\pd_{e}\pd_{f}F_{ac}\pd_{e}\pd_{f}F_{bd}%
  \nn\\&&
  +24\,\g_{ab}\e\, F_{ac}\pd_{d}\pd_{e}F_{bf}\pd_{d}\pd_{f}F_{ce}%
  -16\,\g_{ab}\e\, F_{ac}\pd_{d}\pd_{e}F_{bf}\pd_{d}\pd_{e}F_{cf}%
  \nn\\&&
  -2\,\g_{ab}\e\, F_{ab}\pd_{c}\pd_{d}F_{ef}\pd_{c}\pd_{d}F_{ef}%
  -4\,\g_{abcd}\e\, F_{ae}\pd_{f}\pd_{g}F_{bc}\pd_{f}\pd_{g}F_{de}%
  \nn\\&&
  -4\,\g_{abcd}\e\, F_{ae}\pd_{f}\pd_{g}F_{bc}\pd_{e}\pd_{f}F_{dg}%
  -\,\g_{abcdef}\e\, F_{ab}\pd_{g}\pd_{h}F_{cd}\pd_{g}\pd_{h}F_{ef}%
  \nn\\&&
  -12\,\g_{ab}\e\,\pd_{c}F_{de}\pd_{c}F_{df}\pd_{e}\pd_{f}F_{ab}%
  +8\,\g_{ab}\e\,\pd_{c}F_{de}\pd_{d}F_{cf}\pd_{e}\pd_{f}F_{ab}%
  \nn\\&&
  +64\,\g_{ab}\e\,\pd_{c}F_{de}\pd_{d}F_{af}\pd_{c}\pd_{e}F_{bf}%
  -48\,\g_{ab}\e\,\pd_{c}F_{de}\pd_{f}F_{ad}\pd_{c}\pd_{e}F_{bf}%
  \nn\\&&
  -56\,\g_{ab}\e\,\pd_{c}F_{de}\pd_{f}F_{ad}\pd_{c}\pd_{f}F_{be}%
  +64\,\g_{ab}\e\,\pd_{c}F_{de}\pd_{f}F_{ad}\pd_{e}\pd_{f}F_{bc}%
  \nn\\&&
  -16\,\g_{ab}\e\,\pd_{c}F_{de}\pd_{d}F_{af}\pd_{e}\pd_{f}F_{bc}%
  -8\,\g_{ab}\e\,\pd_{c}F_{de}\pd_{d}F_{af}\pd_{c}\pd_{f}F_{be}%
  \nn\\&&
  -4\,\g_{ab}\e\,\pd_{c}F_{ad}\pd_{b}F_{ef}\pd_{c}\pd_{d}F_{ef}%
  -16\,\g_{ab}\e\,\pd_{c}F_{ad}\pd_{e}F_{bf}\pd_{c}\pd_{f}F_{de}%
  \nn\\&&
  -6\,\g_{ab}\e\,\pd_{c}F_{de}\pd_{f}F_{ab}\pd_{c}\pd_{f}F_{de}%
  +4\,\g_{abcd}\e\,\pd_{e}F_{af}\pd_{f}F_{bg}\pd_{e}\pd_{g}F_{cd}%
  \nn\\&&
  -4\,\g_{abcd}\e\,\pd_{e}F_{af}\pd_{g}F_{bc}\pd_{f}\pd_{g}F_{de}%
  +16\,\g_{ab}\e\,\pd_{c}F_{ad}\pd_{e}F_{bf}\pd_{c}\pd_{e}F_{df}%
  \nn\\&&
  -\,\g_{abcdef}\e\,\pd_{g}F_{ab}\pd_{h}F_{cd}\pd_{g}\pd_{h}F_{ef} \big\}\,.%
\eea

\subsection{Nonlinear supersymmetry\label{nonlintrans}}

The nonlinear supersymmetry transformations of the vector field are:
\bea
  \d A_a &=& a_{(2,0)}\ap{2}\big\{
  +\tfrac{1}{4}F_{ab}\,\bar{\eta}\g_{b}\chi%
  -\tfrac{1}{8}F_{bc}\,\bar{\eta}\g_{abc}\chi \big\}%
  \nn\\&&
  \nn\\&&
 +a_{(2,0)}^2\ap{4}\big\{
  -\tfrac{1}{768}F_{bc}F_{de}F_{fg}\,\bar{\eta}\g_{abcdefg}\chi%
  +\tfrac{1}{128}F_{bc}F_{ad}F_{ef}\,\bar{\eta}\g_{bcdef}\chi%
  \nn\\&&
  +\tfrac{3}{128}F_{bc}F_{bc}F_{de}\,\bar{\eta}\g_{ade}\chi%
  +\tfrac{1}{16}F_{ab}F_{bc}F_{de}\,\bar{\eta}\g_{cde}\chi%
  \nn\\&&
  +\tfrac{1}{32}F_{bc}F_{bd}F_{ce}\,\bar{\eta}\g_{ade}\chi%
  -\tfrac{3}{64}F_{bc}F_{bc}F_{ad}\,\bar{\eta}\g_{d}\chi%
  \nn\\&&
  -\tfrac{1}{16}F_{ab}F_{bc}F_{cd}\,\bar{\eta}\g_{d}\chi \big\}%
  \nn\\&&
  \nn\\&&
  +a_{(4,4)}\ap{4} \big\{
  -20\,\pd_{b}\pd_{c}F_{ad}\,\bar{\eta}\g_{d}\pd_{b}\pd_{c}\chi%
  +12\,\pd_{b}\pd_{c}F_{ad}\,\bar{\eta}\g_{b}\pd_{c}\pd_{d}\chi%
  \nn\\&&
  -4\,\pd_{b}\pd_{c}F_{de}\,\bar{\eta}\g_{ade}\pd_{b}\pd_{c}\chi \big\}\,.%
\eea
The nonlinear supersymmetry transformations of the fermion are:
\bea
  \d \chi &=& \eta
  \nn\\&&
  +a_{(2,0)}\ap{2}\big\{
  +\tfrac{1}{16}F_{ab}F_{ab}\,\eta%
  +\tfrac{1}{32}F_{ab}F_{cd}\,\g_{abcd}\eta \big\}%
  \nn\\&&
  \nn\\&&
 +a_{(2,0)}^2\ap{4} \big\{
  +\tfrac{1}{512}F_{ab}F_{ab}F_{cd}F_{cd}\,\eta%
  -\tfrac{3}{128}F_{ab}F_{ac}F_{bd}F_{cd}\,\eta%
  \nn\\&&
  +\tfrac{1}{512}F_{ab}F_{ab}F_{cd}F_{ef}\,\g_{cdef}\eta%
  -\tfrac{1}{64}F_{ab}F_{ac}F_{bd}F_{ef}\,\g_{cdef}\eta%
  \nn\\&&
  +\tfrac{1}{6144}F_{ab}F_{cd}F_{ef}F_{gh}\,\g_{abcdefgh}\eta \big\}%
  \nn\\&&
  \nn\\&&
  +a_{(4,4)}\ap{4}\big\{
  -4\,\pd_{a}\pd_{b}F_{cd}\pd_{a}\pd_{b}F_{cd}\,\eta%
  +8\,\pd_{a}\pd_{b}F_{cd}\pd_{a}\pd_{b}F_{ce}\,\g_{de}\eta%
  \nn\\&&
  -16\,\pd_{a}\pd_{b}F_{cd}\pd_{a}\pd_{c}F_{be}\,\g_{de}\eta%
  -2\,\pd_{a}\pd_{b}F_{cd}\pd_{a}\pd_{b}F_{ef}\,\g_{cdef}\eta \big\}%
\eea


\begin{thebibliography}{99}
\bibitem{Tseyt1}A.A.~Tseytlin, {\sl Born-Infeld action, supersymmetry
                and string theory},
                Yuri Golfand Memorial Volume, de.\ M.~Shifman,
                World Scientific (2000),
                {\tt hep-th/9908105}
\bibitem{Fradkin}E.S.~Fradkin and A.A.~Tseytlin,
                {\sl Non-linear electrodynamics from quantized strings},
                Phys.~Lett.~{\bf 163B} (1985) 123
\bibitem{APS}  M. Aganagic, C. Popsescu and J.H. Schwarz,
               {\sl Gauge-invariant and Gauge-Fixed D-Brane Actions},
               Nucl. Phys. {\bf B495} (1997) 99-126,
               {\tt hep-th/9612080};
\bibitem{ET}   E. Bergshoeff and P.K. Townsend,
               {\sl Super-D-branes},
               Nucl. Phys. {\bf B490} (1997) 145-162,
               {\tt hep-th/9611173};
\bibitem{CGNSW}M. Cederwall, A. von Gussich, B.E.W. Nilsson, P. Sundell and A. Westerberg,
               {\sl The Dirichlet Super-$p$-Branes in Ten-Dimensional Type IIA and IIB Supergravity},
               Nucl. Phys. {\bf B490} (1997) 179-201,
               {\tt hep-th/9611159}
\bibitem{goteborg} M. Cederwall, B.E.W. Nilsson and D. Tsimpis,
               {\sl The structure of maximally supersymmetric Yang-Mills
               theory: constraining higher-order corrections},
               JHEP 0106 (2001) 034,
               {\tt hep-th/0102009};
               {\sl D=10 super Yang-Mills at $\ap{2}$},
               JHEP 0107 (2001) 042,
               {\tt hep-th/0104236}
\bibitem{KS1}  P.~Koerber and A.~Sevrin,
               {\sl The non-abelian open superstring effective action
               through order $\ap{3}$},
               JHEP 0110 (2001) 003,
               {\tt hep-th/0108169}
\bibitem{RSTZ} A.~Refolli, A.~Santambrogio, N.~Terzi and D.~Zanon,
               {\sl $F^5$  contribution to the non-abelian Born-Infeld action
               from a supersymmetric Yang-Mills five-point function},
               Nucl.~Phys.~{\bf B613} (2001) 64,
               {\tt hep-th/0105277}
\bibitem{Grasso} D.T. Grasso,
               {\sl Higher order contributions to the effective action of $\mathcal{N}=4$
               super-Yang-Mills},
               {\tt hep-th/0210146}
\bibitem{CdRE} A.~Collinucci, M.~de Roo and M.G.C.~Eenink,
               {\sl Supersymmetric Yang-Mills theory at order $\ap{3}$},
               JHEP 06 (2002) 024,
               {\tt hep-th/0205150}
\bibitem{BrMaMe} F.T.~Brandt, F.R.~Machado, R.~Medina,
               {\sl The open superstring 5-point amplitude revisited},
               JHEP 07 (2002) 071,
               {\tt hep-th/0208121}
\bibitem{KS3}  P.~Koerber and A.~Sevrin.
               {\sl The non-abelian D-brane effective action through
               order $\ap{4}$},
               JHEP 10 (2002) 046,
               {\tt hep-th/0208044}
\bibitem{BdRS} E.A.~Bergshoeff, M.~de Roo and A.~Sevrin,
               {\sl Non-abelian Born-Infeld and kappa-symmetry},
               J. Math. Phys. {\bf 42} (2001) 2872,
               {\tt hep-th/0011018}
\bibitem{BBRS} E.A.~Bergshoeff, A.~Bilal, M.~de Roo and A.~Sevrin,
               {\sl Supersymmetric non-abelian Born-Infeld revisited},
               JHEP 0107 (2001) 029,
               {\tt hep-th/0105274}
\bibitem{Sor}  D.~Sorokin,
               {\sl Coincident (Super)-Dp-Branes of Codimension One},
               JHEP 0108 (2001) 022,
               {\tt hep-th/0106212}
\bibitem{DrHoLi} J.M.~Drummond, P.S.~Howe, U.~Lindstr\"{o}m,
               {\sl Kappa-symmetric non-abelian Born-Infeld actions in three dimensions},
               {\tt hep-th/0206148}
\bibitem{PP}   S.S.~Pal and S.~Panda,
               {\sl Coincident Dp-branes in codimension two},
               {\tt hep-th/0211115}
\bibitem{GW}   D.J.~Gross and E.~Witten,
               {\sl Superstring modifications of Einstein's equations},
               Nucl.~Phys.~{\bf B277} (1986) 1
\bibitem{goteborg2} M.~Cederwall, B.E.W.~Nilsson and D.~Tsimpis,
               {\sl Spinorial cohomology of abelian $d=10$
                super-Yang-Mills at $\mathcal{O}(\ap{3})$},
               {\tt hep-th/0205165}
\bibitem{AndTs}O.D.~Andreev and A.A.~Tseytlin,
               {\sl Partition function representation for the open
               superstring effective action: cancellation of M\"obius
               infinities and derivative corrections to
               Born-Infeld Lagrangian},
               Nucl.~Phys.~{\bf B311} (1988) 205
\bibitem{Wyll} N.~Wyllard,
               {\sl Derivative corrections to D-brane actions with constant
                background fields},
               Nucl.~Phys.~{\bf B598} (2001) 247,
               {\tt hep-th/0008125}
\bibitem{Wyll2}N.~Wyllard,
               {\sl Derivative corrections to the D-brane Born-Infeld action:
                non-geodesic embeddings and the Seiberg-Witten map},
               JHEP 0108 (2001) 027,
               {\tt hep-th/0107185}
\bibitem{DMS}  S.R.~Das, S.~Mukhi and N.V.~Suryanarayana,
               {\sl Derivative corrections from noncommutativity},
               JHEP 0108 (2001) 039,
               {\tt hep-th/0106024}
\bibitem{Schwarz}J.H.~Schwarz,
               {\sl Superstring Theory},
               Phys.~Rept.~{\bf 89} (1982) 223
\bibitem{Bilal}A. Bilal, {\sl Higher-derivative corrections to the
               non-abelian Born-Infeld action},
               Nucl.~Phys.~{\bf B618} (2001) 21,
               {\tt hep-th/0106062}
\bibitem{Proey} A.~van Proeyen,
               {\sl Tools for supersymmetry},
               {\tt hep-th/9910030}
\bibitem{BRS}  E.~Bergshoeff, M.~Rakowski and E.~Sezgin,
               {\sl Higher-derivative super Yang-Mills theories},
               Phys.~Lett.~{\bf B185} (1987) 371




\end{thebibliography}
\end{document}